\documentclass[a4paper,11pt]{article}
\usepackage{pos}

\usepackage[english]{babel}
\usepackage{tikz}
\usepackage{graphicx}
\usepackage{wasysym}
\usepackage{wrapfig}
\usepackage{listings}
\usepackage[absolute,verbose,overlay]{textpos}
\usepackage{amsmath}
\usepackage{bbold}
\usepackage{bm}
\usepackage{braket}
\usepackage{colortbl}
\usepackage{hhline}
\usepackage{multirow}
\usepackage{animate}
\usepackage{xfrac}

\usepackage{lineno}
\usepackage{accents}
\usepackage{acronym}

\acrodef{BCF}[BCF]{Borici-Creutz fermions}
\acrodef{BC}[BC]{Borici-Creutz}
\acrodef{KSF}[KSF]{Kogut-Susskind fermions}
\acrodef{KS}[KS]{Kogut-Susskind}
\acrodef{KWF}[KWF]{Karsten-Wilczek fermions}
\acrodef{KW}[KW]{Karsten-Wilczek}
\acrodef{MDF}[MDF]{Minimally doubled fermions}
\acrodef{ST}[ST]{spin-taste}

\definecolor{hu-berlin-blue}{RGB}{0,65,137} 
\definecolor{hu-berlin-green}{RGB}{150,190,20} 
\definecolor{hu-berlin-grey}{RGB}{169,169,169}
\definecolor{hu-berlin-brown}{RGB}{82,79,60}
\definecolor{hu-berlin-red}{RGB}{180,0,0}
\definecolor{wsu-green}{RGB}{0,89,76}
\definecolor{mc-gill-red}{RGB}{237, 27, 47}
\definecolor{tum-blue}{RGB}{0,101,189}
\definecolor{clustergray}{RGB}{117,128,145}
\definecolor{clusterorange}{RGB}{255,158,37}

\newcommand{\ee}{\textrm{e}} 
\newcommand{\ii}{\textrm{i}} 

\newcommand{\bmat}{\left(\begin{array}}
\newcommand{\emat}{\end{array}\right)}
\newcommand{\bvec}{\left(\begin{array}{c}}
\newcommand{\evec}{\end{array}\right)}

\renewcommand{\dag}{^\dagger}

\newcommand{\abs}[1]{|{#1}|}

\newcommand{\vr}{\varrho}

\newcommand{\bdm}{\begin{displaymath}}
\newcommand{\edm}{\end{displaymath}}

\newcommand{\si}{\sigma}
\newcommand{\ga}{\gamma}

\newcommand{\Ga}{\Gamma}

\newcommand{\Dnai}{D_\text{nai}}

\newcommand{\SKW}{S_\text{KW}}
\newcommand{\SBC}{S_\text{BC}}

\newcommand{\mC}{\mathcal{C}}

\newcommand{\mO}{\mathcal{O}}
\newcommand{\mP}{\mathcal{P}}

\newcommand{\mR}{\mathcal{R}}

\newcommand{\mT}{\mathcal{T}}

\renewcommand{\sfrac}[2]{\ensuremath{#1/#2}}

\newcommand{\str}[1]{} 

\newcommand{\pa}[1]{(#1)}

\title{Spin-taste structure of minimally doubled fermions}

\author*[a]{Johannes H. Weber}

\affiliation[a]{Humboldt Universit{\"a}t zu Berlin \& RTG2575,\\
  Zum Gro{\ss}en Windkanal 2, 12489 Berlin, Germany}

\abstract{
Minimally doubled fermions realize one degenerate pair of Dirac fermions on the lattice. 
Similarities to staggered fermions exist, namely, spin and taste degrees of freedom become 
intertwined, and a remnant, non-singlet chiral symmetry and ultralocality are maintained. 
However, charge conjugation, isotropy and some space-time reflection symmetries are broken 
by the cutoff.
For two variants, i.e., Karsten-Wilczek (KW) or Borici-Creutz (BC) fermions, a tasted charge 
conjugation symmetry can be identified, and the respective representations of the spin-taste 
algebra can be constructed explicitly. In the case of BC fermions, the tasted symmetry 
indicates that amendments to the published counterterms are necessary.
The spin-taste representation on the quark level permits construction of local or extended 
hadron interpolating operators for any spin-taste combination, albeit with contamination by 
parity partners and taste-symmetry violation. The few available numerical results for KW 
fermions are in line with expectations.\\
\\
HU-EP-23/68-RTG
}

\FullConference{The 40th International Symposium on Lattice Field Theory (Lattice 2023)\\
July 31st - August 4th, 2023\\
Fermi National Accelerator Laboratory\\}

\begin{document}
\maketitle

\section{Introduction}

\ac{MDF} realize a specific taste-vector chiral symmetry at finite lattice cutoff with a strictly local operator and the minimal number of Dirac fermions, namely two, for arbitrary even dimension $D$ in compliance with the Nielsen-Ninomiya No-Go theorem~\cite{Nielsen:1980rz}.
On the one hand, the \ac{KW} variant, which has its tastes located on one axis of the Brillouin zone, has been proposed in the early 80s~\cite{Karsten:1981gd, Wilczek:1987kw}. 
On the other hand, the \ac{BC} variant has its tastes located on a hypercubic diagonal and has been proposed in the late 2000s~\cite{Creutz:2007af, Borici:2007kz}. 
Both \ac{BC} or \ac{KW} variants have been shown (for $D=2$) to perceive the gauge field topology correctly~\cite{Durr:2022mnz} and yield renormalizable QFTs that require analogous patterns of counterterms, which are known in perturbation theory~\cite{Capitani:2009yn, Capitani:2010nn}, or--in some cases--non-perturbatively~\cite{Weber:2013tfa, Weber:2015hib}\footnote{Non-perturbative tuning of dynamical \ac{KW} fermions or in the valence sector has been discussed at this conference.}. 
Further variants going by the names of twisted-ordering or dropped twisted-ordering exist~\cite{Creutz:2010cz}, but become indistinguishable from \ac{BC} or \ac{KW} variants for $D=2$~\cite{Durr:2020yqa}, respectively. 

The construction of \ac{MDF} is similar to Wilson fermions insofar as operators of lower mass dimension are added to the naive fermion operator.
At the tree-level, these cancel at their leading mass dimension for a subset of tastes that play the role of surviving Dirac fermions with opposite chirality in the continuum limit. 
The necessary fine-tuning in the interacting theory gives rise to the relevant counterterm. 
For \ac{MDF} such extra operators are $\ga_{5}$ hermitian products of $\gamma$ matrices and discretized Laplacian components, times an $r$ parameter (some restrictions apply to realize minimal doubling). 
Thus, symmetries under charge conjugation and some subset of space-time reflections are broken, while the symmetry under the product $\mC\mP\mT$ remains intact. 
For the \ac{KW} variant one may preserve the symmetry under either time reflection or parity~\cite{Pernici:1994yj}, while neither can be preserved for the \ac{BC} variant~\cite{Creutz:2008sr}. 
In order to understand the \ac{ST} structure, one might try \ac{ST} diagonalization that would reduce naive fermions to $2^{D/2}$ identical copies of \ac{KS} fermions~\cite{Kluberg-Stern:1983lmr}. 
However, this does not produce a diagonal subgroup when applied to \ac{BC} or \ac{KW} variants, but instead identical \ac{ST} structures in the two separate taste $U(2^{D/2})$ groups~\cite{Kimura:2011ik}. 
Since the \ac{ST} structures differ between the single-site and site-split contributions to the Laplacian terms, \ac{ST} diagonalization is generally impossible for \ac{MDF}. 
While ad-hoc site-splitting prescriptions could approximately filter the two tastes, e.g. as suggested for \ac{KW} fermions~\cite{Creutz:2010qm} or \ac{BC} fermions~\cite{Basak:2017oup}, they cannot accommodate the two different non-trivial \ac{ST} structures in each of the two variants, and cannot bring forth the exact \ac{ST} representations of $\mathfrak{su}(2)$ for these discretizations. 
In the interacting theory the taste symmetry is approximate, and the eigenvalues arrange into taste multiplets~\cite{Ammer:2022ksl}.

Fortunately, on the one hand, the \ac{ST} structure of the \ac{KW} variant is straightforward to derive from first principles, and emerges naturally in the chiral representation~\cite{Weber:2016dgo}. 
In Sec.~\ref{sec:KW}, we revisit the \ac{KW} representation of the taste $\mathfrak{su}(2)$ algebra, and derive a \ac{ST} basis of meson-interpolating operators as our first new result. 
We relate these to published results on spin-zero meson correlation functions~\cite{Weber:2015oqf}. 
Unfortunately, on the other hand, the \ac{ST} structure of the \ac{BC} variant is not straightforward at all. 
Building on the same ideas we can construct a \ac{ST} representation for the \ac{BC} variant, too, if the corresponding $r$ parameter is appropriately restricted. 
This derivation is our second new result and discussed in Sec.~\ref{sec:BC}, where we correct the published result for the marginal fermionic counterterm, our third new result. 
We close in Sec.~\ref{sec:summary} with a comment on extensions of these \ac{ST} representations to different $r$ or to twisted-ordering operators, and an outlook towards phenomenologically relevant applications of \ac{MDF}.

\section{Karsten-Wilczek fermions}\label{sec:KW}

The standard \ac{KW} action (in the conventions and notation of~\cite{Durr:2020yqa}) reads
\begin{equation}
    \SKW[{{\psi}}, \bar{{\psi}}] 
    = 
    a^D \sum\limits_{n,m \in {\Lambda}}
    \bar{{\psi}}\pa{n} 
    \Big[
    \Dnai[U] ~+~m_0
    ~
    -\frac{ra}{2} \ii \ga_{D}
    \sum\limits_{j=1}^{D-1} \Delta_{j}[U]
    \Big](n,m)~
    {\psi}\pa{m}
    \label{eq:SKW}~,
\end{equation}
\noindent
where we define the naive Dirac operator and discretized Laplacian (components) with gauge-covariant translation operators $t_{\pm\mu}[U](n,m) \equiv U_{\pm\mu}(n)\delta(n \pm\hat{\mu},m)$, 
$U_{-\mu}(n) \equiv U_{\mu}\dag(n-\hat{\mu})$ 
as 

\begin{align}
    \Dnai[U](n,m)
    &\equiv\sum\limits_{\mu=1}^{D} \ga_{\mu}
    \frac{s_{+\mu}[U](n,m)}{a}
    =\sum\limits_{\mu=1}^{D} \ga_{\mu}
    \frac{t_{+\mu}[U](n,m)-t_{-\mu}[U](n,m)}{2a}
    \label{eq:nai}~,\\
    \Delta_{\mu}[U](n,m) 
    &\equiv 
    \frac{2c_{\mu}[U](n,m)-2\delta(n,m)}{a^2}
    = 
    \frac{t_{+\mu}[U](n,m)+t_{-\mu}[U](n,m)-2\delta(n,m)}{a^2}
    \label{eq:Lap}~.
\end{align}

\noindent
The \ac{KW} action as described in Eq.~\eqref{eq:SKW} is obviously invariant under discrete translations (for periodic boundary conditions), discrete spatial rotation-reflections ($W_3$, and in particular, parity $\mP$), and the product of (Euclidean) time reflection ($\mT$) and charge conjugation ($\mC$). 
Furthermore, it is $\ga_{5}$ hermitian and satisfies a chiral symmetry with (unmodified) $\ga_5$ for $m_0 \to 0$. 
Anisotropy between temporal ($D$) and spatial ($1 \le j \le D-1$) directions is a consequence of the individually broken symmetry under $\mT$ that maps the $r$-parameter as $r \to -r$. 
For $r^2 > \sfrac{1}{4}$, the \ac{KW} action is minimally doubled with the survivors located at $ak_D=0,\pi$ and $ak_j=0$ in the Brillouin zone. 
Thus, it seems natural to define a \ac{KW} fermion hyper-site as two boson sites one step apart in the ${D}$-direction. 
Necessarily, a translation $t_{\pm D}[U](n,m)$ within that hyper-site is a \ac{ST} vector transform; 
thus, any translation by an odd number of steps in the ${D}$-direction must be a \ac{ST} vector transform, too. 

The \ac{KW} action inherits the $\mu=D$ mirror fermion symmetry~\cite{Pernici:1994yj} of the naive operator in Eq.~\eqref{eq:nai}, which is a symmetry under $ak_\mu \to \pi -ak_\mu$ in momentum space.
In position space this is understood as being due to the product of a (Euclidean) reflection of the $\mu$-direction ($\mR_{\mu}$), where $\mR_{D} = \mT$ is just (Euclidean) time reflection, and a \ac{ST} rotation ($\tau_{\mu} \equiv \tau_{\mu}(n)=\tau_{\mu}\dag(n)= \ga_{\mu5} (-1)^{n_\mu}$, with hermitian matrices $\ga_{\mu\nu} \equiv \ii \ga_{\mu} \ga_{\nu}$ for $1 \le \mu < \nu \le 5$)

\begin{align}
    \mR_{\mu} \psi\pa{n_\nu,n_\mu}  &= \ga_{\mu} \ga_5 \psi\pa{n_\nu,-n_\mu} 
    ~,&~
    \bar{\psi}\pa{n_\nu,n_\mu} \mR_{\mu}\dag  &= \bar{\psi}\pa{n_\nu,-n_{\mu}}  \ga_5 \ga_{\mu}
    \label{eq:Rmu}~,\\
    \tau_{\mu}(n) \psi\pa{n_\nu,n_\mu}  &= \ga_{\mu5} (-1)^{n_\mu} \psi\pa{n_\nu,n_\mu} 
    ~,&~
    \bar{\psi}\pa{n_\nu,n_\mu} \tau_{\mu}\dag(n)  &= \bar{\psi}\pa{n_\nu,n_\mu} (-1)^{n_\mu} \ga_{\mu5}
    \label{eq:taumu}~,
\end{align}

\noindent
where we have omitted the standard transformation of the gauge fields under $\mR_{\mu}$; see e.g.~\cite{Gattringer:2010zz}. 
Obviously, this mirror fermion symmetry (under $\tau_{D}\mT$) implies that the \ac{ST} rotation $\tau_{D}$ flips the sign of the $r$ parameter, too, such that the \ac{KW} action has a \ac{ST} charge conjugation symmetry.
There is a \ac{KW} representation of the generators $\{\rho_i(n)\} \equiv \{\rho[\si_i] \}$ of the $\mathfrak{su}(2)$ algebra, namely $[\rho_i,\rho_j] =2\ii \epsilon_{ijk} \rho_{k}$, yielding well-defined transformation patterns for all parts of Eq.~\eqref{eq:SKW}, commuting with the continuum limit of $\Dnai[U](n,m)$, and with $\tau_{D}(n)$ as a generator. 
Specifically\footnote{In the chiral representation of the Euclidean gamma matrices $\rho_{i}(n) \propto \sigma_{i} \otimes \mathbb{1}$ suggest a natural identification.}
\begin{equation}
\begin{split}
    \rho_{1}(n) &= \ga_{D} (-1)^{p_S(n)}
    ~,\\
    \rho_{2}(n) &= \ii\ga_{D} \ga_{5} (-1)^{p_D(n)} \equiv \tau_{D}(n)
    ~,\\
    \rho_{3}(n) &= \ga_{5} (-1)^{p(n)} = \ga_{5} \epsilon(n) \equiv \tau_{5}(n)
    ~,
\end{split}
    \label{eq:su2KW}
\end{equation}
\noindent
where we have introduced site parities restricted to spatial, temporal, or all components ($p_S(n)=(n_1+\ldots+n_{D-1})\mod 2,~p_D(n)=n_D\mod 2, p(n)=(n_1+\ldots+n_{D})\mod 2$). 
The staggered $\epsilon(n)=(-1)^{p(n)}$ maps between survivors at $ak_j=0$ and lifted doublers at $ak_j=\pi$.
Site-split or extended operators with translations transform non-trivially under the $\rho_i(n)$, such that we conclude 
\begin{equation}
\begin{split}
    t_{\pm D}[U](n,m)  &\sim \rho_{1}(n,m)
    ~,\\
    t_{\pm j}[U](n,m) &\sim \rho_{2}(n,m)
    ~,\\
    t_{\pm j}[U](n,l)t_{\pm D}[U](l,m) &\sim \rho_{3}(n,m)
    ~,    
\end{split}
    \label{eq:splitKW}
\end{equation}
where the symbol ``$\sim$'' means that site-split path transforms under $\rho_{j}(n)$ as $\rho_{j}(n)\rho_{i}(n,m)\rho_{j}(m) = (-1)^{\delta_{ij}-1}\rho_{i}(n,m)$. 
While an even number of steps in any one direction is a \ac{ST} singlet transform for \ac{KW} fermions, 
naive site-splitting procedures~\cite{Creutz:2010qm} fail to implement the \ac{ST} structure, since the $\rho_{i}(n,m)$ do not satisfy the $\mathfrak{su}(2)$ algebra.
Because translations in different directions do not commute in the interacting theory, 
$\rho_{\mu}(n,l)\rho_{\nu}(l,m) \neq \rho_{\nu}(n,l)\rho_{\mu}(l,m)$, we see that a \ac{ST} vector mass term $\sim \rho_{3}(n,m)$ must renormalize differently from a \ac{ST} singlet mass term. 

\begin{table}[ht]
    \centering
    \begin{tabular}{c||c|c|c|c|c|c|c|c}
        $\otimes$       &  $\Ga_{5}$ 
                        &  $\Ga_{D}$
                        &  $\Ga_{j}$
                        &  $\Ga_{jk}$
                        &  $\Ga_{jD}$
                        &  $\Ga_{j5}$
                        &  $\Ga_{D5}$
                        &  $\mathbb{1}$ 
        \\
        \hline
        \hline
        {$\rho_{1}$}    & ${(\ga_{D})}$ 
                        & $\ga_{D}$ 
                        & ${[\ga_{j5} \epsilon]}$ 
                        & ${(\ga_{j})}$ 
                        & ${(\ga_{j5})}$ 
                        & $\ga_{j5}$ 
                        & ${[\ga_{D} \epsilon]}$ 
                        & --
        \\
        {$\rho_{2}$}    & -- 
                        & ${[\ga_{D5} \epsilon]}$ 
                        & $\ga_{j}$
                        & --
                        & --
                        & ${[\ga_{j} \epsilon]}$ 
                        & $\ga_{D5}$ 
                        & ${(\ga_{D5})}$ 
        \\
        {$\rho_{3}$}    & $\ga_{5}$ 
                        & ${(\ga_{5})}$ 
                        & ${(\ga_{jk})}$ 
                        & ${[\ga_{jD} \epsilon]}$ 
                        & $\ga_{jD}$
                        & ${(\ga_{jD})}$ 
                        & --
                        &  ${[\ga_{5} \epsilon]}$ 
        \\
        {$\mathbb{1}$}  & ${[\epsilon]}$ 
                        & --
                        & --
                        & $\ga_{jk}$
                        & ${[\ga_{jk} \epsilon]}$ 
                        & --
                        & ${(\mathbb{1})}$ 
                        &  $\mathbb{1}$ 
        \\
        \hline
        \hline
        {$\rho_{1}$}    & $c_{j}\ga_{5}$ 
                        & ${(c_{j}\ga_{5})}$  
                        & ${(c_{D}\ga_{jk})}$ 
                        & $c_{D}\ga_{jk}$
                        & $c_{j}\ga_{jD}$
                        & ${(c_{j}\ga_{jD})}$ 
                        & ${(c_{D}\mathbb{1})}$ 
                        & $c_{D}\mathbb{1}$
        \\
        {$\rho_{2}$}    & $c_{D}\ga_{5}$ 
                        & ${(c_{D}\ga_{5})}$ 
                        & ${(c_{j}\ga_{jk})}$ 
                        & $c_{j}\ga_{jk}$
                        & $c_{D}\ga_{jD}$
                        & ${(c_{D}\ga_{jD})}$ 
                        & ${(c_{j}\mathbb{1})}$ 
                        & $c_{j}\mathbb{1}$
        \\
        {$\rho_{3}$}    & ${(c_{j}\ga_{D})}$ 
                        & $c_{j}\ga_{D}$ 
                        & $c_{D}\ga_{j}$ 
                        & ${(c_{D}\ga_{j})}$ 
                        & ${(c_{j}\ga_{j5})}$ 
                        & $c_{j}\ga_{j5}$ 
                        & $c_{D}\ga_{D5}$ 
                        & ${(c_{D}\ga_{D5})}$ 
        \\
        {$\mathbb{1}$}  & ${(c_{D}\ga_{D})}$
                        & $c_{D}\ga_{D}$ 
                        & $c_{j}\ga_{j}$ 
                        & ${(c_{j}\ga_{j})}$ 
                        & ${(c_{D}\ga_{j5})}$ 
                        & $c_{D}\ga_{j5}$ 
                        & $c_{j}\ga_{D5}$ 
                        & ${(c_{j}\ga_{D5})}$  
        \\
    \end{tabular}%
    \caption{Spin-taste assignment of meson-interpolating operators for \ac{KW} fermions. 
    Columns/rows identify spin($\Gamma$)/taste($\rho$) structures; entries indicate $M(n,m)$ of $\bar{\psi}\pa{n} M(n,m)\psi\pa{m}$.
    $M$ represents continuum/Wilson fermion spin assignment, while $(M)$ indicates the assignment for the parity partner. 
    Top: Single-site/zero-link operators. 
    $[M]$ indicates a state of desired quantum numbers at the three-momentum cutoff; two-link operators $c_j c_D M$ achieve the same assignment near zero momentum.
    Bottom: One-link operators $c_{\mu} M$.
    }
    \label{tab:mesonKW}
\end{table}

When we split the Laplacian according to Eq.~\eqref{eq:Lap}, the different \ac{ST} structures of the terms in the \ac{KW} action emerge, which are the same as in Table 1 of Ref.~\cite{Weber:2016dgo}. 
Since odd powers in the cutoff are combined with odd powers in $r$,
conclusions regarding automatic cancellation of odd powers of $r$ and $a$ in the \ac{KW} determinant can be taken over\footnote{This is a stronger statement than automatic $\mO(a)$ improvement, since it applies to any odd powers.}. 
Thus, one may average both signs of $r$ in the valence sector and enforce explicit cancellation of any odd powers in the valence sector without extra fine tuning. 
For meson-interpolating operators, we only need to classify between zero- or one-link operators, shown in Table~\ref{tab:mesonKW}, since two-link operators can be related---up to the major subtlety, which Fermi points are connected---to the zero-link ones. 
Symmetrization is optional, but reduces noise, i.e.  $t_{\pm\mu}$ instead of $c_{\mu}$ yields the same \ac{ST} structures.

We studied spin-zero single-site operators on pure gauge backgrounds~\cite{Weber:2015oqf} and found for naive fermions perfect degeneracy between the respective parity partners $M=\ga_5$ vs $\ga_{D}$ or $M=\ga_{D5}$ vs $\mathbb{1}$. For \ac{KW} fermions, however, there is no degeneracy between $M=\ga_{5}$ or $\ga_{D}$ (while $M=\ga_{D5}$ vs $\mathbb{1}$ remains inconclusive due to large statistical errors). 
The splitting between $M=\ga_{5}$ vs $\ga_{5D}$ is similar for both. 
Thus, two different sources of taste-symmetry violation affect the \ac{KW} variant.

\section{Borici-Creutz fermions}\label{sec:BC}

The standard \ac{BC} action (in the conventions of~\cite{Durr:2020yqa}) and in a convenient notation reads

\begin{equation}
    \SBC[{{\psi}}, \bar{{\psi}}] 
    = 
    a^D \sum\limits_{n,m \in {\Lambda}}
    \bar{{\psi}}\pa{n} 
    \Big[
    \Dnai[U] ~+~m_0
    -\frac{r a}{2} \sum\limits_{\mu=1}^{D}
    \ii \ga_{\mu}^{\prime} c_{\mu}[U]
    +\frac{r}{a} \ii \Ga
    \Big](n,m)~
    {\psi}\pa{m}~,
    \label{eq:SBC}
\end{equation}
\noindent
The dual gamma matrices $\ga_{\mu}^{\prime}$ are generated by selecting one hypercubic-diagonal direction via $\Ga$,
\begin{equation}
    \ga_{\mu}^{\prime} = \Ga \ga_{\mu} \Ga
    ~,\quad
    \Ga \equiv \frac{1}{\sqrt{d}} \sum\limits_{\mu=1}^{d}\ga_{\mu}
    = \frac{1}{\sqrt{d}} \sum\limits_{\mu=1}^{d}\ga_{\mu}^{\prime}
    ~.
    \label{eq:dualBC}
\end{equation}
The \ac{BC} action as described in Eqs.~\eqref{eq:SBC} and~\eqref{eq:dualBC} is obviously invariant under discrete translations (for periodic boundary conditions), 
discrete rotations around the hypercubic diagonal or axis exchanges (leaving $\Ga$ invariant), and the product $\mC\mP\mT$. 
Furthermore, it is $\ga_{5}$ hermitian and satisfies a chiral symmetry with (unmodified) $\ga_5$ for $m_0 \to 0$. 
Anisotropy between the hypercubic-diagonal direction and those orthogonal to it is a consequence of the broken symmetry under $\mP\mT$ (or $\mC$) that maps the $r$-parameter as $r \to -r$. 
For $r^2 > \sfrac{1}{2}$ (in $D=4$), the \ac{BC} action is minimally doubled with the survivors located at $ak_\mu=0$ or $-2\arctan(\sfrac{1}{r})$ (all $k_{\mu}$ equal) in the Brillouin zone. 
The forward/backward symmetry along the hypercubic-diagonal direction is broken; cf. Eq.~\ref{eq:dualBC}.

In order to identify the \ac{BC} representation of taste $\mathfrak{su}(2)$ we first define a unitary \ac{ST} rotation \vskip-3ex

\begin{align}
    &
    \psi\pa{n} \stackrel{T_{\Ga}}{\to} T_{\Ga}(n) ~\psi\pa{n}
    ~,~\quad~
    \bar{\psi}\pa{n} \stackrel{T_{\Ga}}{\to} \bar{\psi}\pa{n}~T_{\Ga}^{\dagger}(n) 
    \label{eq:TGapsi}~,\\
    &
    T_{\Ga}(n) \equiv \ii \Ga \ga_{5} \xi_{r}^{q(n)}
    \hskip10pt,\quad 
    \xi_{r} \equiv \ii s_{r} \equiv \ii~\text{sign}(r)
    \hskip11pt,\quad 
    q(n) = (n_1+\ldots+n_{D})\mod 4
    \label{eq:TGadef}~,\\
    &
    T_{\Ga}(n) ~\ga_{\mu}~ T_{\Ga}\dag(m) = -\ga_{\mu}^{\prime} \xi_{r}^{q(n-m)}
    ~\Rightarrow\quad
    T_{\Ga}(n) ~\Ga~ T_{\Ga}\dag(m) = -\Ga \xi_{r}^{q(n-m)}
    ~,
\end{align}

\noindent
which converts Eq.~\eqref{eq:SBC} into
\begin{equation}
    \SBC[{{\psi}}, \bar{{\psi}}] 
    = 
    a^D \sum\limits_{n,m \in {\Lambda}}
    \bar{{\psi}}\pa{n} 
    \Big[
    \abs{r}
    \Dnai[U] ~+~m_0
    ~
    +\frac{s_{r} a}{2} \sum\limits_{\mu=1}^{D}
    \ii \ga_{\mu}^{\prime} c_{\mu}[U]
    -\frac{r}{a} \ii \Ga
    \Big](n,m)~
    {\psi}\pa{m}~
    \label{eq:SBCtrans}
\end{equation}
in terms of the transformed fields. Note that an odd power in $\abs{r}$ is swapped between the derivative and Laplacian terms. 
For $r^2=1$ (and thus $\abs{r}=1, s_r=r$)--to which we restrict hereafter--the form of Eq.~\eqref{eq:SBC} is restored with $r \to -r$ under $\mC$ (or $\mP\mT$). Thus, we have identified an additional \ac{ST} charge conjugation symmetry seemingly similar to the one of the \ac{KW} variant. 
A certain awkwardness follows from the unitary nature of $T_{\Ga}(n)$--hermitian/antihermitian for even/odd $n$-- cf. Eq.~\eqref{eq:TGadef}. 
The block decomposition of $\Ga$, 
\begin{equation}
    \Ga 
    = \bmat{cc} 0 & R \\ R^{\dagger} & 0\emat 
    ~,\quad
    R = \sqrt{\frac{2}{d}} 
    \bmat{cc}   \vr^{-1} & \vr^{-3} \\ 
                \vr^{-1} & \vr^{+1} \emat
    ~,\quad
    \vr=\frac{1+\ii}{\sqrt{2}} 
    = \ee^{\frac{\ii\pi}{4}}
    ~,
    \label{eq:blockBC}
\end{equation}
permits us to write down a \ac{BC} representation of the generators of taste $\mathfrak{su}(2)$ in all their glory,
\begin{equation}
\begin{split}
        {
        \rho_1(n) 
        }&{
        = \phantom{\ii}
        \bmat{cc} 0 & (-)^{p(n)\phantom{+1}}R \\ R^{\dagger} & 0\emat 
        ~\xi_r^{(2p(n)-1)q(n)} 
        \phantom{^{+}}
        = \phantom{\ii} \Ga\ga_{5}^{p(n)}
        \phantom{^{+1}}
        ~\xi_r^{(2p(n)-1)q(n)} 
        \phantom{^{+}}
        = [-\ii T_{\Ga} T_{5}^{p(n)-1} ](n)
        }~,\\
        {
        \rho_2(n) 
        }&{
        = \ii 
        \bmat{cc} 0 & (-)^{p(n)+1}R \\ R^{\dagger} & 0\emat
        ~\xi_r^{(2p(n)+1)q(n)} 
        \phantom{^{-}}
        = \ii \Ga \ga_{5}^{p(n)+1} 
        ~\xi_r^{(2p(n)+1)q(n)} 
        \phantom{^{-}}
        = [T_{\Ga} T_{5}^{p(n)} ](n)
        }~,\\
        {
        \rho_3(n) 
        }&{
        = \phantom{\ii}
        \bmat{cc} \mathbb{1}\phantom{^{\dagger}} & 0 \\ 0 & 
        \phantom{^{(1+}}-\mathbb{1}\phantom{^{p(n))}} \emat 
        ~\xi_r^{2q(n)}
        \phantom{^{(1+-p(n))}}
        = \phantom{\ii}\phantom{\Ga}\ga_{5}\phantom{^{(n)+1}}
        ~\xi_r^{2q(n)} 
        \phantom{^{(1+-p(n))}}
        \equiv T_{5}(n)
        }~,
\end{split}
        \label{eq:su2BC}
\end{equation}
which indeed satisfy the usual $\mathfrak{su}(2)$ algebra $[\rho_i,\rho_j] =2\ii \epsilon_{ijk} \rho_{k}$ for arbitrary $n$.
We note that $\rho_{3}(n)$ is the same real operator as in Eq.~\eqref{eq:su2KW} for \ac{KW} fermions, i.e. $T_{5}(n) = \tau_{5}(n)$, since $\xi_r^{2q(n)}=\epsilon(n)$. 
All terms in Eq.~\eqref{eq:SBC} have well-defined transformation behaviors under the generators in Eq.~\eqref{eq:su2BC}: 
on the one hand, the site-split ones acquire extra factors of $(-\ga_{5})=\rho_{3}(\pm\hat{\mu})$, while the taste-vector single-site one acquires only powers of $(-1)$. 
On the other hand, the combinations $T_{\Ga}(n)$--which we used to obtain Eq.~\eqref{eq:SBCtrans}--(or $-\ii[T_{\Ga}T_{5}](n)$) do not introduce factors of $\ga_{5}$. 
The unitary operators $T_{\Ga}(n)$ (or $-\ii[T_{\Ga}T_{5}](n)$) correspond to $\rho_{2/1}(n)$ (or $\rho_{1/2}(n)$, respectively) for even/odd sites (up to phases), such that the generators in the \ac{ST} charge conjugation symmetry combine to one power of $\rho_{3}(n)$ for both site-split terms (up to phases). 

With the taste generators of Eq.~\eqref{eq:su2BC} in hand, it is straightforward to write down two-link operators that transform non-trivially under the $\rho_i(n)$ (or $T_{\Ga}(n)$ or $-\ii[T_{\Ga}T_{5}](n)$, respectively), such that we conclude (for arbitrary $1 \le \mu,\nu \le D$)
\begin{equation}
    t_{\pm(\mu+\nu)}[U](n,m) \sim \rho_{3}(n,m)
    ~,\quad
    t_{\pm(\mu-\nu)}[U](n,m) \sim \mathbb{1}\delta(n,m)
    ~.    
    \label{eq:two-linkBC}
\end{equation}
Thus, any forward/backward hops from even/odd sites have the same \ac{ST} structure, and combine with any forward/backward hops from odd/even sites to $\rho_{3}(n,m)$ (independent of $\mu,\nu$). 
Most importantly, we can write down a \ac{ST} vector mass term via Eq.~\eqref{eq:two-linkBC}, which commutes both with $\mP\mT$ or $\mC$ such that there is a simultaneous eigenbasis. 
Each hop individually contains a factor of the chirality matrix $\ga_{5}$ and a forward/backward dependent phase factor $\xi_{r}^{\pm1}$: 
\begin{equation}
\begin{split}
    \rho_{1}(n) t_{\pm\mu}[U](n,m) \rho_{1}(m) \sim - \ga_{5} \xi_{r}^{\pm1} t_{\pm\mu}[U](n,m)
    ~,\\
    \rho_{2}(n) t_{\pm\mu}[U](n,m) \rho_{2}(m) \sim + \ga_{5} \xi_{r}^{\pm1} t_{\pm\mu}[U](n,m)
    ~.    
\end{split}
    \label{eq:one-linkBC}
\end{equation}
Eq.~\eqref{eq:two-linkBC} follows trivially from Eq.~\eqref{eq:one-linkBC}. 
However, the one-link operators in Eq.~\eqref{eq:one-linkBC} do not permit a simultaneous eigenbasis with $\mP\mT$ or $\mC$, and one-link site-splitting~\cite{Basak:2017oup} does not access the tastes. 

\begin{table}[ht]
    \centering
    \begin{tabular}{c||c|c|c|c|c||c|c}
                        & {$\Dnai$}  
                        & {$\ii r\ga_{\mu}^{\prime}{c}_\mu$}
                        & {$\ii r\Ga$} 
                        & {$\mathbb{1}$} 
                        & {$c_{\mu+\nu}$} 
                        & {$\Ga {s}_\mu$}
                        & {$\ii r\Ga {c}_\mu$}
        \\
        \hline
        \hline
        $\mP\mT$        & + & - & - & + & + & + & - \\
        $\mathcal{C}$   & + & - & - & + & + & + & - \\
        {
        $-\ii[T_{\Ga}T_{5}]$}     & {$+\ii s_r \ga_{\mu}^{\prime}{c}_\mu$} 
                        & {$-\abs{r}\Dnai$}  & + & + & - 
                        & {$+\ii s_r \Ga {c}_\mu$} & {$- \abs{r} \Ga {s}_\mu$} \\
        {
        $T_{\Ga}$}      & {$+\ii s_r \ga_{\mu}^{\prime}{c}_\mu$} 
                        & {$-\abs{r}\Dnai$} & - & + & - 
                        & {$+\ii s_r \Ga {c}_\mu$} & {$- \abs{r} \Ga {s}_\mu$} \\
        {
        $T_{5}$}        & + & + & - & + & + & + & + \\
        $\ga_{5}$       & - & - & - & + & + & + & + \\
    \end{tabular}%
    \caption{Spin-taste structure of the renormalized \ac{BC} action 
    including the marginal fermionic counterterm.
    }
    \label{tab:patternBC}
\end{table}

We collect the \ac{ST} structure of all fermionic terms of the renormalized \ac{BC} action in Table~\ref{tab:patternBC}, and see yet another important consequence of the \ac{ST} charge conjugation symmetry.
The 2nd to last column is the marginal fermionic counterterm as suggested in Ref.~\cite{Capitani:2009yn}, which is mapped onto itself by $\mP\mT$ or $\mC$, yet mapped onto the last column by $T_{\Ga}(n)$ (or $-\ii[T_{\Ga}T_{5}](n)$).
Thus, the correct counterterm combines both columns, propped up by the single-site term, as 
\begin{equation}
    a^D \sum\limits_{n,m \in {\Lambda}}
    \bar{{\psi}}\pa{n} 
    \Ga \sum_{\mu=1}^{D}
    \Big[ s_\mu[U] -\frac{\ii r}{a} \big(c_{\mu}[U]-1\big) \Big](n,m) ~
    {\psi}\pa{m}
    ~,
    \label{eq:rightBC}
\end{equation}
where the $c_{\mu}[U](n,m)$ structure makes the counterterm compliant with the \ac{ST} charge conjugation symmetry, while the constant cancels the relevant contribution due to the $c_{\mu}[U](n,m)$ structure. 
Eq.~\eqref{eq:rightBC} supersedes Ref.~\cite{Capitani:2009yn}. 
Attempts to tune with the wrong counterterm produced explicit $\mT$-symmetry violation in pion correlators for any non-zero coefficient~\cite{Basak:2017oup}, 
while a correct counterterm respecting the symmetries as in Eq.~\eqref{eq:rightBC} will not induce it despite mistuning.

Ultimately, it is the different taste structures of the site-split and the single-site terms in the 
\ac{BC} action that are responsible for automatic cancellation of odd powers of $r$ and $a$ in the \ac{BC} determinant. 
The argument is based on symmetry under $\mC$, $T_{5}$ and $\ga_{5}$, analogous to the one for the \ac{KW} variant~\cite{Weber:2016dgo}, and does not depend on the condition $r^2=1$. 
Automatic cancellation of odd powers is possible in the valence sector similar to the \ac{KW} variant, too. 
However, since parity partners appear as complex oscillating contributions (with period $4a$) in any directions that have non-zero projection on the hypercubic diagonal (cf. Eq.~\ref{eq:dualBC}), identification of the \ac{ST} structures is much more complicated in the case of the \ac{BC} variant. 
A similar clarity as in Table~\ref{tab:mesonKW} cannot be achieved unless the hypercubic diagonal is used as the time direction. 
Because there is no real incentive, we do not compose a similar table here. 
Zero- or two-link operators could be worked out from Eq.~\eqref{eq:two-linkBC}, while 
one-link operators need substantially more care than naively applying Eq.~\eqref{eq:one-linkBC}.

\section{Summary}\label{sec:summary}

We have clarified the spin-taste structure of \ac{BC} and \ac{KW} variants. 
Studying the \ac{KW} variant is far easier~\cite{Weber:2016dgo}, and we classified zero-, one-, and two-link meson interpolating operators taking into account the parity partner contributions. 
We also derived the \ac{ST} structure for the \ac{BC} variant relying on an \ac{ST} charge conjugation symmetry for $r^2=1$. 
We used this symmetry to correct the marginal fermionic counterterm~\cite{Capitani:2009yn}.
While we expect that the \ac{BC} representation of the taste $\mathfrak{su}(2)$ algebra and the form of the counterterm hold for general $r$ or for twisted-ordering operators, there is no substitute for the \ac{ST} charge conjugation symmetry. 
Yet the properties of the determinant and automatic cancellation of odd powers in $a$ can be generalized.
We advise to avoid \ac{BC} or twisted-ordering variants with their extreme complexities and drawbacks in favor of the \ac{KW} variant.

\section*{Acknowledgements}

J.H.W.’s research is funded by the Deutsche Forschungsgemeinschaft (DFG, German Research
Foundation)---Projektnummer 417533893/GRK2575 ``Rethinking Quantum Field Theory''. 
J.H.W. thanks Stephan D{\"u}rr for fruitful discussions and their careful reading of the manuscript.

\end{document}